\documentclass[a4paper]{jpconf}

\usepackage[utf8]{inputenc}
\usepackage[T1]{fontenc}
\usepackage[english]{babel}
\usepackage[numbers,square,sort&compress]{natbib}
\usepackage{graphicx}
\usepackage{xcolor}
\usepackage{mathtools}
\usepackage{amssymb}
\usepackage{lmodern}
\usepackage{siunitx}
\usepackage[final]{microtype}
\usepackage{hyperref}

\hypersetup{%
	pdftitle = {Dyson-Schwinger approach to baryon number fluctuations},
	pdfauthor = {Philipp Isserstedt, Michael Buballa, Christian S. Fischer, Pascal J. Gunkel},
	bookmarksopen = true,
	colorlinks = true,
	linkcolor = black,
	citecolor = black,
	urlcolor = black
}

\graphicspath{{./Graphics/}}

\newcommand*{\+}{\hspace*{.08335em}}
\newcommand*{\beq}{\begin{equation}}
\newcommand*{\eeq}{\end{equation}}
\newcommand*{\Nf}{N_{\textup{f}}}
\newcommand*{\Nc}{N_{\textup{c}}}
\newcommand*{\Tc}{T_{\textup{c}}}
\newcommand*{\dd}{\textup{d}}
\newcommand*{\upB}{\textup{B}}
\newcommand*{\upS}{\textup{S}}
\newcommand*{\upQ}{\textup{Q}}
\newcommand*{\upu}{\textup{u}}
\newcommand*{\upd}{\textup{d}}
\newcommand*{\ups}{\textup{s}}
\newcommand*{\muu}{\mu_{\upu}}
\newcommand*{\mud}{\mu_{\upd}}
\newcommand*{\mus}{\mu_{\ups}}
\newcommand*{\muB}{\mu_{\upB}}
\newcommand*{\muS}{\mu_{\upS}}
\newcommand*{\muQ}{\mu_{\upQ}}
\newcommand*{\bbN}{\mathbb{N}}
\newcommand*{\bbZ}{\mathbb{Z}}
\newcommand*{\srs}{\sqrt{s}}
\providecommand*{\Tr}{\operatorname{Tr}}
\DeclarePairedDelimiterX{\expval}[1]{\langle}{\rangle}{#1}

\DeclareSIUnit{\MeV}{\mega\electronvolt}
\DeclareSIUnit{\MeV}{\giga\electronvolt}

\begin{document}

\title{Dyson-Schwinger approach to baryon number fluctuations}

\author{%
	Philipp Isserstedt$^1$,
	Michael Buballa$^2$,
	Christian S Fischer$^1$
	and
	Pascal J Gunkel$^1$
}

\address{%
	$^1$ Institut f\"{u}r Theoretische Physik,
	Justus-Liebig-Universit\"{a}t Gie\ss{}en,
	35392 Gie\ss{}en,
	Germany
}

\address{%
	$^2$ Theoriezentrum,
	Institut f\"{u}r Kernphysik,
	Technische Universit\"{a}t Darmstadt,
	64289 Darmstadt,
	Germany
}

\ead{philipp.isserstedt@physik.uni-giessen.de}

\begin{abstract}
We summarize our results on baryon number fluctuations at nonzero temperature and chemical potential. They are obtained from solutions of a coupled set of Dyson-Schwinger equations for the quark and gluon propagators of QCD in Landau gauge with $\Nf=2+1$ quark flavors. In comparison with preliminary STAR data, our results are compatible with a critical endpoint at large chemical potential and a freeze-out line that bends below it.
\end{abstract}

\section{\label{sec:intro}Introduction}

Proving the very existence and possibly locating the theoretically conjectured critical endpoint (CEP)
in the phase diagram of QCD is one of the main quests of contemporary heavy-ion collision experiments.
Assuming that the freeze-out happens close to the chiral crossover line, fluctuations of conserved charges
(baryon number, strangeness, and electric charge) are expected to be promising quantities to provide
signals of the CEP in experiments. Ratios of cumulants of these conserved quantities can be obtained in
event-by-event analyses and compared to ratios obtained in theoretical calculations. See, e.g.,
Refs.~\citep{Asakawa:2015ybt,Luo:2017faz,Bzdak:2019pkr} for review articles.

In the following we summarize our recent results \citep{Isserstedt:2019pgx} on baryon number fluctuations
at nonzero temperature $T$ and baryon chemical potential $\muB$ as well as our updated result for the
QCD phase diagram with $\Nf=2+1$ physical quark flavors using the nonperturbative framework of Dyson-Schwinger
equations (DSEs).

\section{\label{sec:fluctuations}Fluctuations}

Fluctuations of the conserved quantities baryon number (B), strangeness (S), and electric charge (Q)
in heavy-ion collisions are given by derivatives of QCD's grand-canonical potential $\Omega$ with
respect to the corresponding chemical potentials, viz.
\beq
	\label{eq:fluctuations}
	\chi_{lmn}^{\upB\upS\upQ}
	=
	-\frac{1}{T^{\+ 4-(l+m+n)}}
	\frac{\partial^{\+ l+m+n} \, \Omega}{\partial \muB^l \+ \partial \muS^m \+ \partial \muQ^n}
\eeq
with $l, m, n \in \bbN$. The connection to experiment is established, for example, by the ratios
\beq
	\chi_3^\upB / \chi_2^\upB = S_\upB \+ \sigma_\upB
	\, , \qquad
	\chi_4^\upB / \chi_2^\upB = \kappa_\upB \+ \sigma_\upB^2
\eeq
of baryon number fluctuations with $S_\upB$, $\sigma_\upB$, and $\kappa_\upB$ denoting the skewness,
variance, and kurtosis of the net-baryon distribution, respectively. Analogous expressions hold for
strangeness and electric charge. Results from the Beam Energy Scan program at the Relativistic Heavy
Ion Collider at Brookhaven National Laboratory for fluctuations of the net-proton number obtained by
the STAR collaboration \citep{Luo:2015ewa,Luo:2015doi} can be used as a proxy for fluctuations of the
net-baryon number and therefore serve for comparisons.

We use the nonperturbative framework of DSEs to provide results for fluctuations of the baryon
number in ($2+1$)-flavor QCD at nonvanishing $T$ and $\muB$. They are determined via the quark
number densities
\beq
	\label{eq:density}
	\rho_f
	=
	-\frac{\partial \+ \Omega}{\partial \mu_f}
	=
	-\Nc \+ Z_2^f \+ T \, \sum_{n \in \bbZ} \, \int \+ \frac{\dd^3 \+ \vec{p}}{(2\pi)^3}
	\Tr \bigl[ \gamma_4 \+ S_f(p) \bigr] \, ,
\eeq
where $f \in \{ \upu, \upd, \ups \}$, $\Nc = 3$ denotes the number of colors, $Z_2^f$ is the quark wave
function renormalization constant, and $p = (\omega_n, \vec{p}\,)$ with Matsubara frequencies
$\omega_n = (2n+1) \+ \pi T$; $n \in \bbZ$. The quark chemical potentials $\mu_f$ are related to
the ones for baryon number, strangeness and electric charge via
$\muu = \muB \+ / \+ 3 + 2 \+ \muQ \+ / \+ 3$,
$\mud = \muB \+ / \+ 3 - \muQ \+ / \+ 3$, and
$\mus = \muB \+ / \+ 3 - \muQ \+ / \+ 3 - \muS$.
Using these relations, one expresses the derivatives appearing in Eq.~\eqref{eq:fluctuations} in terms
of Eq.~\eqref{eq:density} and derivatives thereof. More details can be found in
Ref.~\citep{Isserstedt:2019pgx}. The quantity $S_f$ denotes the dressed
(i.e., nonperturbative) quark propagator at nonzero temperature and chemical potential and will be
detailed in the next section.

\section{\label{sec:dse}Dyson-Schwinger equations}

\begin{figure}[t]
	\centering%
	\includegraphics[trim=0 -5.8mm 0 0]{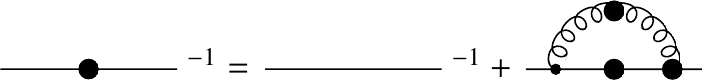}%
	\hfill%
	\includegraphics{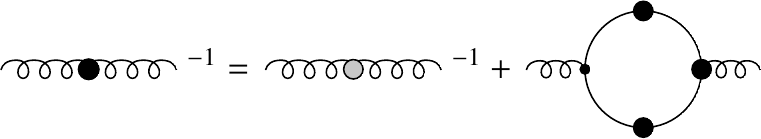}%
	\caption{\label{fig:dse}%
		DSEs for the quark (left) and gluon (right) propagators. Large filled circles denote dressed
		quantities; solid and wiggly lines represent quarks and gluons, respectively. There is a
		separate DSE for each quark flavor. In the gluon equation, the gray circle denotes the bare
		gluon propagator together with all diagrams with no explicit quark content and the quark-loop
		diagram contains an implicit flavor sum. The diagrams were drawn with JaxoDraw \citep{Binosi:2008ig}.
	}
\end{figure}

The dressed quark propagator at nonzero temperature and chemical potential is needed to
compute the quark number densities, Eq.~\eqref{eq:density}. It is obtained by solving a coupled set of
truncated DSEs where the backcoupling of quarks onto the gluon is taken explicitly into account.
This allows for a consistent mass and flavor dependence of the gluon beyond simple models. Furthermore,
the gluon becomes sensitive to the chiral dynamics of the quark.
The dressed quark and gluon propagators each satisfy a DSE shown diagrammatically in Fig.~\ref{fig:dse}.
They are coupled to higher-order correlation functions, e.g., the dressed quark-gluon vertex,
which obey their own DSEs. Thus, truncations are necessary to obtain a closed system of equations.
We replace all pure Yang-Mills self-energies in the gluon DSE, i.e., all diagrams with no explicit
quark content, with quenched temperature-dependent lattice data \citep{Fischer:2010fx,Maas:2011ez}
and add the quark loop explicitly. For the dressed quark-gluon vertex we use the following ansatz:
The leading tensor structure of the Ball-Chiu vertex construction \citep{Ball:1980ay}, which solves the
Abelian Ward-Takahashi identity for the vertex, is supplemented with a phenomenological infrared-enhanced
function.

For the sake of brevity, we do not show explicit expressions here and refer the reader to
Ref.~\citep{Isserstedt:2019pgx} for more details. This setup, which gives us access to the dressed
quark and unquenched gluon propagators at arbitrary temperature and chemical potential, has been
successfully used in a series of previous works. See Ref.~\citep{Fischer:2018sdj} and references therein
for a comprehensive overview.

\section{\label{sec:results}Results and discussion}

\subsection{Phase diagram}

\begin{figure}[t]
	\centering%
	\includegraphics[trim=0 2mm 0 0]{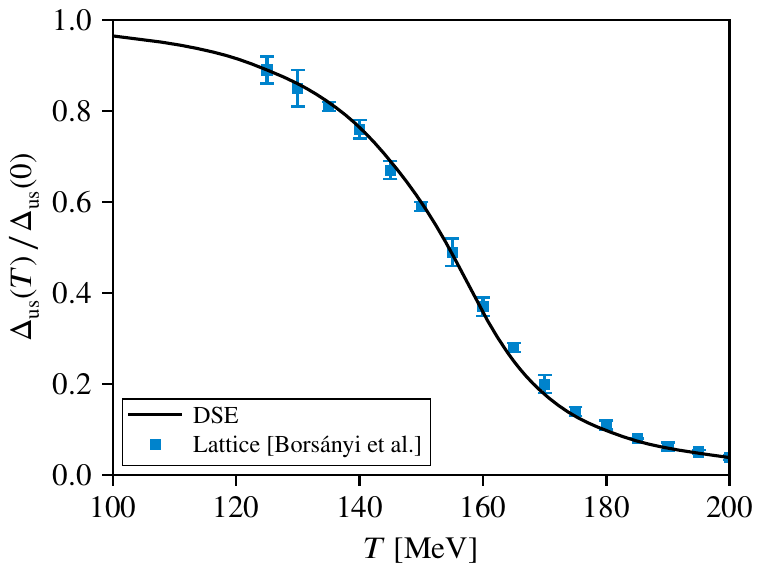}%
	\hfill%
	\includegraphics[trim=0 2mm 0 0]{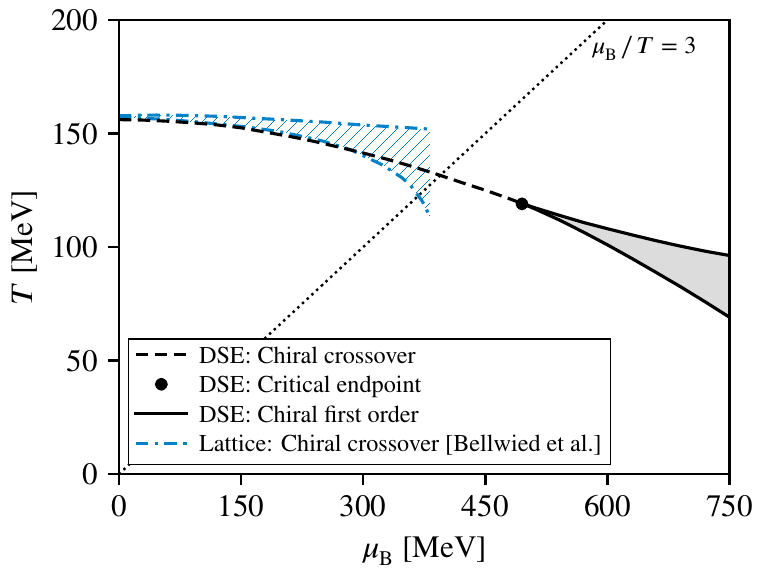}%
	\caption{\label{fig:condensate_phase_diagram}%
		Left:
		Normalized subtracted condensate at vanishing chemical potential compared to
		continuum-extrapolated lattice results from Ref.~\citep{Borsanyi:2010bp}.
		Right:
		Our result for the QCD phase diagram with $\Nf=2+1$ physical quark flavors. The blue
		band is the region of the chiral crossover from lattice QCD \citep{Bellwied:2015rza}.
		Figures slightly modified taken from Ref.~\citep{Isserstedt:2019pgx}.
	}
\end{figure}

The order parameter for chiral symmetry breaking is the quark condensate
\beq
	\expval{\bar{\psi}\psi}_f
	=
	- \Nc \+ Z_2^f Z_m^f \+ T \, \sum_{n \in \bbZ} \, \int \+
	\frac{\dd^3 \+ \vec{p}}{(2\pi)^3} \Tr\bigl[ S_f(p) \bigr] \, ,
\eeq
where $Z_m^f$ is the mass renormalization constant. The quark condensate is divergent for nonzero current-quark
masses $m_{\upu,\ups}$ and needs to be regularized. This is done by considering the so-called subtracted
condensate
$\Delta_{\upu\ups} = \expval{\bar{\psi} \psi}_\upu - (m_\upu \+ / \+ m_\ups) \+ \expval{\bar{\psi} \psi}_\ups$.
We use the inflection point of $\Delta_{\upu\ups}$ with temperature to define the pseudocritical chiral
transition temperature $\Tc$. The infrared strength of our vertex ansatz is chosen such that $\Tc$ obtained
in this way coincides with lattice results. Again, more details can be found in Ref.~\citep{Isserstedt:2019pgx}.

In the left diagram of Fig.~\ref{fig:condensate_phase_diagram} we show our result for the subtracted
condensate (solid black) at vanishing chemical potential compared to continuum-extrapolated lattice
results from Ref.~\citep{Borsanyi:2010bp} (blue squares).
As described above, our result of $\Tc = \SI{156}{\MeV}$ agrees by construction with the result from
lattice QCD. A nontrivial result is, however, the almost perfect match of the steepness.
The right diagram of Fig.~\ref{fig:condensate_phase_diagram} shows our result for the QCD phase diagram
with $\Nf=2+1$ quark flavors at the physical point. The crossover line (dashed black) becomes steeper
with increasing chemical potential and finally terminates in a second-order CEP at
$\bigl( T^{\+\textup{CEP}}, \, \muB^\textup{CEP} \bigr) = (119, 495) \, \si{\MeV}$
followed by the coexistence region of a first-order transition (shaded gray) bound by spinodal lines
(solid black). The CEP occurs at large chemical potential with a corresponding ratio of
$T^{\+\textup{CEP}} / \+ \muB^\textup{CEP} \approx 4.2$. We also show results for the chiral crossover
from lattice QCD \citep{Bellwied:2015rza} (blue band) obtained by analytic continuation from imaginary
to real chemical potential and find good agreement with our crossover line. 

\subsection{Baryon number fluctuations}

\begin{figure}[t]
	\centering%
	\includegraphics[trim=0 2mm 0 0]{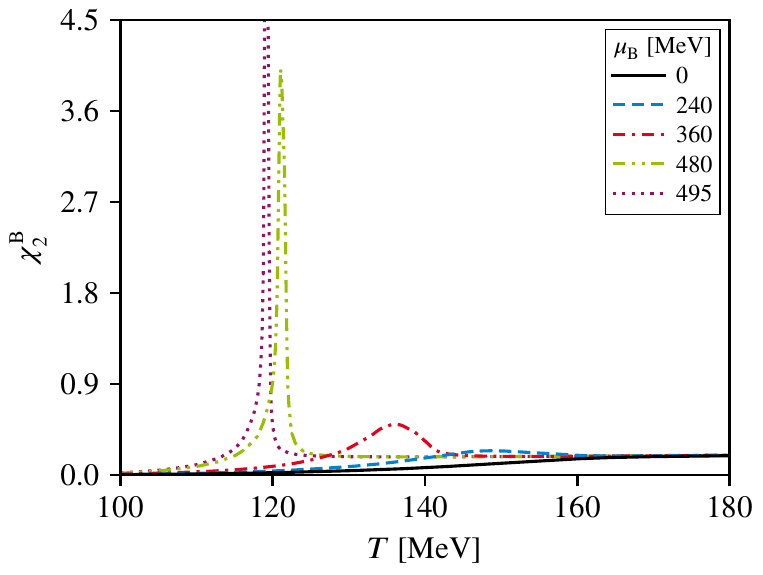}%
	\hfill%
	\includegraphics[trim=0 2mm 0 0]{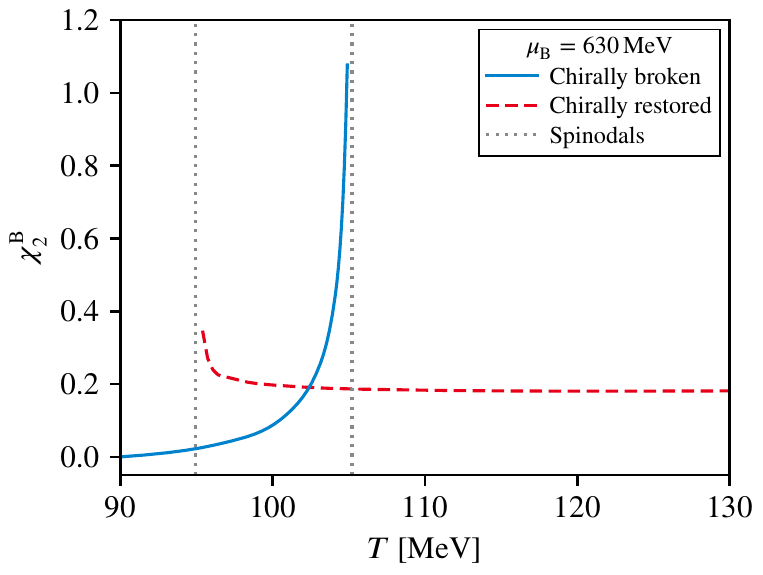}%
	\caption{\label{fig:fluctuations_01}%
		Second-order baryon number fluctuation approaching the CEP (left) and in the first-order
		region of the phase diagram (right). Figures taken from Ref.~\citep{Isserstedt:2019pgx}.
	}
\end{figure}

In the left diagram of Fig.~\ref{fig:fluctuations_01} we show the second-order baryon number fluctuation
$\chi_2^\upB$ approaching the CEP. At vanishing chemical potential (solid black) we find a monotonous
increase with the maxium of the slope located near $\Tc$. At nonzero chemical potential, a bulge
begins to form around the pseudocritical chiral transition temperature and grows with increasing
chemical potential (dashed blue and dash-dotted red). In the vicinity of the CEP, the bulge becomes
a sharp peak with a significant rise in magnitude (dash-dot-dotted green) and finally diverges at
the CEP (dotted purple). This is indeed expected since at the CEP the correlation length $\xi$ of
the system diverges (at least for infinite volume), $\xi \to \infty$, and $\chi_2^\upB \sim \xi^c$
with $c > 0$.

The behavior of $\chi_2^\upB$ beyond the CEP in the first-order transition region of the phase diagram
is displayed in the right diagram of Fig.~\ref{fig:fluctuations_01}. The second-order baryon number
fluctuation splits into two branches corresponding to two different solutions of the DSE for the
dressed quark propagator: the chirally broken solution (solid blue) and the partially chirally restored
solution (dashed red). The region where both solutions are found defines the coexistence region of
the first-order transition bounded by spinodals (vertical dotted gray lines). Outside this coexistence
region, $\chi_2^\upB$ is only a slowly varying function with temperature.

\begin{figure}[t]
	\centering%
	\includegraphics[trim=0 2mm 0 0]{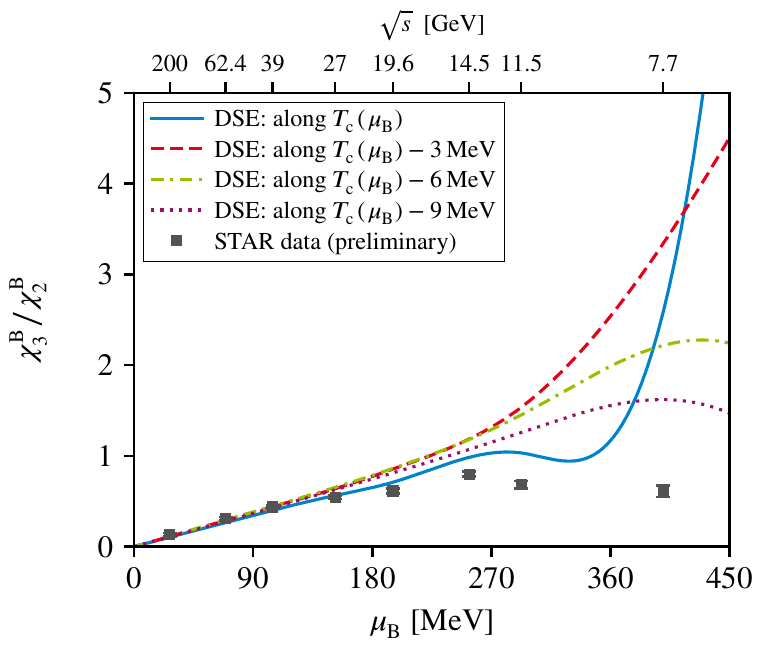}%
	\hfill%
	\includegraphics[trim=0 2mm 0 0]{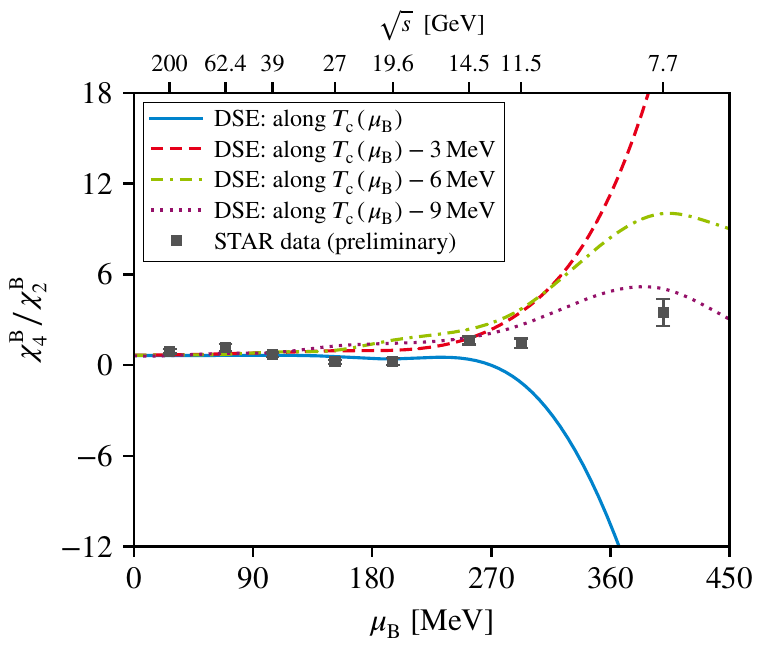}%
	\caption{\label{fig:fluctuations_02}%
		Skewness ratio $\chi_3^\upB / \chi_2^\upB$ (left) and kurtosis ratio
		$\chi_4^\upB / \chi_2^\upB$ (right) along the crossover line and along
		lines with a fixed distance in temperature from it in comparison to preliminary
		data from the STAR collaboration \citep{Luo:2015ewa,Luo:2015doi} at most central
		collisions. We adopt the $\muB$-$\srs$ translation from Ref.~\citep{Adamczyk:2017iwn}.
		Figures taken from Ref.~\citep{Isserstedt:2019pgx}.
	}
\end{figure}

Even though one has to be cautious when comparing theoretical results for fluctuations with experimental
data, there is considerable interest to do so.%
\footnote{The caveats involved are related to the experimental situation in heavy-ion collisions.
For example finite-volume effects, the finite temporal extent of the fireball, and whether/when the system
is in thermal equilibrium. See Refs.~\citep{Luo:2017faz,Bzdak:2019pkr} and references therein.}
In Fig.~\ref{fig:fluctuations_02} we present results for the skewness and kurtosis ratios
$\chi_3^\upB / \chi_2^\upB$ (left) and $\chi_4^\upB / \chi_2^\upB$ (right), respectively, along our
chiral crossover line (solid blue). At small chemical potential (respectively large $\srs\+$) we find
good agreement with data from the STAR collaboration. From $\srs = \SI{14.5}{\GeV}$ on the
agreement becomes worse and disappears for $\srs \leq \SI{11.5}{\GeV}$ since the fluctuations start to
react strongly to the presence of our CEP. We also evaluated these ratios along lines with a constant
difference in temperature of 3, 6, and \SI{9}{\MeV} below the crossover line. The reason is that it is
not clear if the freeze-out and crossover lines have the same curvature. It may be the case that the
freeze-out line bends stronger than the crossover line with increasing chemical potential. This notion is
at least qualitatively supported by our results. At small chemical potential, the variation in both ratios
with increasing temperature distance from the crossover line can only hardly discriminated by the data.
However, the data points at $\srs = \SI{19.6}{\GeV}$ and $\srs = \SI{14.5}{\GeV}$ support a freeze-out
line located very close to the crossover line. We conclude that this is generally the case for
$\srs > \SI{14.5}{\GeV}$. On the other hand, results for the kurtosis ratio at $\srs = \SI{11.5}{\GeV}$
and $\srs = \SI{7.7}{\GeV}$ indicate that the freeze-out and crossover lines are separated by at
least \SI{9}{\MeV} in this region of the phase diagram. The skewness ratio follows the same trend,
although on a less quantitative level than the kurtosis ratio.

Finally, we would like to mention that our results are accompanied with limitations. There may be a
substantial error associated with the precise location of the CEP. It stems entirely from the truncation
of the quark-gluon vertex and may be reduced in the future by more extensive DSE studies
\citep{Eichmann:2015kfa,Contant:2018zpi} and/or systematic comparisons with similar calculations within the
framework of the functional renormalization group \citep{Braun:2009gm,Fu:2019hdw}.
Furthermore, at the current level of truncation, the critical exponents of our CEP are mean field.
We aim to include explicit pion and sigma-meson effects in future work which are expected to put our CEP
in the correct $\textup{Z}(2)$ universality class. 

\ack

This work has been supported by HGS-HIRe for FAIR, the GSI Helmholtzzentrum f\"{u}r Schwerionenforschung,
HIC for FAIR within the LOEWE program of the State of Hesse, the BMBF under contract 05P18RGFCA, and
the Deutsche Forschungsgemeinschaft (DFG, German Research Foundation) -- project number 315477589 -- TRR 211.

\bibliographystyle{iopart-num}
\bibliography{Isserstedt_Fairness}

\end{document}